\documentclass[twoside]{dis04}
\usepackage{amsmath,amssymb}
\newcommand{\pom}{{\mathbb P}}

\def\runauthor{S. Munier and A. Shoshi}
\def\shorttitle{Diffractive photon dissociation in the saturation regime}
\begin{document}

\title{
Diffractive photon dissociation\\
  in the saturation regime
}

\author{S. Munier$^{\dagger}$ and A. Shoshi$^{\ddagger}$
\thanks{supported by the Deutsche Forschungsgemeinschaft under 
contract Sh~92/1-1.}}

\address{
$^\dagger$ Centre de physique th\'eorique (UMR 7644 du CNRS), 
\'Ecole Polytechnique,\\ 91191 Palaiseau cedex, France.\\
$^\ddagger$ 
Department of Physics,
Columbia University, New York, NY 10027, USA.\\
E-mail: munier@cpht.polytechnique.fr, shoshi@phys.columbia.edu }

\maketitle

\abstracts{
Using the Good and Walker picture, we derive
a simple formula for diffractive dissociation that can apply to
recent data collected at HERA in the low $Q^2$ regime.
}

Deep inelastic scattering of a photon of virtuality $Q$
off a given target can be viewed as
the photon splitting into a quark-antiquark pair
of size $r\sim 1/Q$, and subsequently scattering
off the target through a further quantum fluctuation.
This interpretation is valid in the leading logarithmic
approximation of QCD when the total rapidity $Y$ 
is very large,
and in a frame in which $Y$ is mostly carried by
the projectile photon.
The rapidity of the target in that frame is $Y_0\ll Y$.

The initial $q\bar q$ pair is a color dipole. 
The higher Fock states built up from QCD radiation
can also be interpreted as collections of dipoles \cite{Mueller:1993rr}. 
That holds in the large $N_c$ limit
in which a gluon is essentially 
a zero-size $q\bar q$ state, as far as color is concerned.

There are typically two classes of events observed in experiments:
either the target breaks up and the whole rapidity
range is filled with decay products,
or it remains essentially intact and an angular
region empty of final state
particles is seen
in the detector. The latter events are called
diffractive, and may represent up to 20\% of all events. 
Among them, the ones that exhibit a final state
made of the $q\bar q$ pair
together with additional gluonic radiation 
in the photon fragmentation region
are called {\it dissociative}. They are characterized 
experimentally by a large invariant mass $M_X\gg Q$ of the decay products
of the photon.

The main goal of the work on which we report here \cite{Munier:2003zb}
is to provide a simple derivation of the dissociative cross section
valid also in the saturation regime, when effects due to unitarity corrections
become sizeable, and to compare it to recent HERA data.
We also argue that diffractive and total cross sections have
the same rapidity dependence, and this comes about very naturally in
our framework.


\section{High energy behavior of deep-inelastic scattering observables}

From a simple quantum mechanical calculation, one obtains that
the total, elastic and dissociative cross sections 
for the scattering of a color dipole of size $r$
at fixed impact parameter
$b$ are given by
\begin{equation}
\begin{split}
\frac{d\sigma_{tot}}{d^2b}=&2\left(1-\langle S(r,b)\rangle_{Y-Y_0}\right),\ \
 \frac{d\sigma_{el}}{d^2b}=\langle 1- S(r,b)\rangle_{Y-Y_0}^2\\
\frac{d\sigma_{diss}}{d^2b}=&\langle S^2(r,b)\rangle_{Y-Y_0}
-\langle S(r,Y)\rangle_{Y-Y_0}^2\ .
\end{split}
\label{cs}
\end{equation}
The diffractive cross section is the sum of the elastic and dissociative 
ones.
$S(r,b)$ is the $S$-matrix element for the elastic scattering of a
fixed partonic configuration of the dipole of size $r$ at impact parameter $b$.
It is a random variable, whose distribution is related
to the probability distribution of a given Fock state of the 
initial $q\bar q$ pair.

The average $\langle .\rangle_{Y-Y_0}$
is taken over all possible 
partonic configurations of the initial dipole after evolution
over a rapidity range $Y-Y_0$. 
The last formula above
is due to Good and Walker \cite{Good:1960ba}, 
and is an identity between
the dissociative cross section and the variance of the $S$-matrix
for partonic states. The total cross section instead 
is related to the expectation value of the latter.

The evolution law for $S$ with $Y$ can be readily derived.
Within the rapidity interval $dy$, a dipole of size $r$ may
split into two dipoles of size $z$, $r-z$
by emitting
a gluon. That occurs with a probability proportional to $dy$,
say $\lambda\, dy$. Let us denote by $\rho(z,r)$ the distribution of 
the sizes $z$, $r-z$ of the emitted dipoles. Then
\begin{equation}
S(r)\underset{Y\rightarrow Y+dy}\longrightarrow
\left\{
\begin{aligned}
S(r)\phantom{aaa}&\ \ \mbox{with probability}\ 1-\lambda\, dy\\
S(z)S(r-z)&\ \ \mbox{with probability}\ \lambda\, dy\ ,
\end{aligned}
\right.
\end{equation}
$z$ being distributed according to $\rho(z,r)d^2 z$.
That evolution law implies a recursion relation for $\langle S(r)\rangle_Y$:
\begin{equation}
\langle S(r)\rangle_{Y+dy}
=(1-\lambda\, dy)\langle S(r)\rangle_Y
+\lambda\, dy\int d^2 z\, \rho(z,r)
\langle S(z)S(r-z)\rangle_Y\ .
\end{equation}
The limit $dy\rightarrow 0$ yields
\begin{equation}
\partial_Y \langle S(r)\rangle_Y
=\frac{\bar\alpha}{2\pi}\int d^2 z \frac{r^2}{z^2(r-z)^2}
\left(\langle S(r)\rangle_Y -
\langle S(z)S(r-z)\rangle_Y\right)\ .
\label{b}
\end{equation}
We have used the known QCD result for
the dipole splitting probability \cite{Mueller:1993rr}
\begin{equation}
dy\,\lambda\times\rho(z,r)\,d^2z=
dy\frac{\bar\alpha}{2\pi}\frac{r^2}{z^2(r-z)^2}d^2z\ ,
\end{equation}
where $\bar\alpha=\alpha_s N_c/\pi$.
The latter equation is not closed: it 
involves the correlator
$\langle S(z)S(r-z)\rangle_Y$. It turns out to be the first equation
of an infinite hierarchy named after
Balitski\u{\i}. A mean field approximation
$\langle S(z)S(r-z)\rangle\simeq \langle S(z)\rangle\langle S(r-z)\rangle$
casts Eq.(\ref{b}) into a 
closed form, known as the Balitski\u{\i}-Kovchegov (BK) 
equation \cite{Balitsky:1996ub}.

Note that the fact that $S$ is a scattering
matrix element plays no special role in the above derivation.
The only assumption
is that
the created dipoles are independent in such a way that their 
interaction factorizes.
Thus we see that 
the quantity $\langle S^2(r)\rangle_Y$ obeys exactly the
same evolution equation~(\ref{b}), with $S$ replaced by $S^2$.

The BK equation falls in the universality class of the
Fisher-Kolmogorov equation, and was shown \cite{Munier:2003vc} 
to admit traveling
wave solutions at large $Y$, that decay like 
$1-\langle S(r)\rangle_Y
\sim 1-\langle S^2(r)\rangle_Y
\sim (rQ_s(Y))^{2\gamma_0}$
for $r\ll 1/Q_s^2(Y)$ ($\gamma_0\sim 0.63$ is fixed).
$Q_s(Y)$ is the saturation scale.
From Eq.(\ref{cs}), it results that the ratio 
\begin{equation}
\frac{\sigma_{diff}}{\sigma_{tot}}=
\frac{1-2\langle S(r)\rangle_{Y-Y_0}+\langle S^2(r)\rangle_{Y-Y_0}}
{2(1-\langle S(r)\rangle_{Y-Y_0})}\ ,
\end{equation}
which involves quantities that all obey the same equation,
is independent of $Y$. A similar result was obtained by a direct
calculation in Ref.\cite{kl}.

The independence of that ratio with respect to the center-of-mass energy
was seen experimentally in the HERA data. 
However, it seems to hold in rather narrow
bins of the invariant diffracted mass. Here, we have shown that the
total diffractive cross section (with a minimum rapidity gap $Y_0$)
is independent of $Y$, which is a weaker result: it just means that at very high
energy, deep inelastic scattering off a proton and off a Pomeron have the same
energy dependence.


\section{High mass diffraction}

We compute the diffractive cross section to order $\alpha_s$,
allowing for at most one gluon in the diffractive system. Beyond that
one-fluctuation, the $S$-matrix is approximated by its average
$S(z)\sim\langle S(z)\rangle$.
Replacing Eq.(\ref{b}) for $\langle S\rangle$ and $\langle S^2\rangle$ resp.
into Eq.(\ref{cs}) and setting $Y_0=Y_\pom$ the size of the rapidity
gap,
one gets \cite{Kovchegov:2001ni,Munier:2003zb}
\begin{multline}
\frac{d\sigma_{diss}}{d^2b\, d\log(1/x_g)}=\frac{\bar\alpha}{2\pi}
\int d^2z\frac{r^2}{z^2(r-z)^2}\\
\times
({\mathcal S}(Y_\pom,z,b-(r\!-\!z)/2){\mathcal S}(Y_\pom,r-z,b\!-\!z/2)
-{\mathcal S}(Y_\pom,r,b))^2
\label{int}
\end{multline}
where $x_g$ is the longitudinal momentum fraction carried by the
diffracted gluon. It is related to the rapidity
variables through $\log(1/x_g)=Y-Y_\pom$, and to 
the diffracted mass by
$M_X=Q/\sqrt{x_g}$.
${\mathcal S}$ is the dipole elastic
$S$-matrix element in the mean field approximation.

To arrive at HERA phenomenology, one has to convolute Eq.(\ref{int}) with
the distribution of dipoles of size $r$ in the virtual photon, given by the squared wave
function $|\psi(z_q,r)|^2$ for the splitting $\gamma^*\rightarrow q\bar q$ ($x_g$
is the longitudinal momentum fraction carried by the quark). We also assume
the Golec-Biernat-W\"usthoff (GBW) model \cite{GBW}
for ${\mathcal S}$, which in particular has no $b$-dependence.
After some algebra, one gets
\begin{multline}
\frac{d\sigma_{diss}}{dM_X}=\frac{2\alpha_s N_c}{\pi^2}
\frac{\sigma_0}{M_X}
\int d^2 r \int_0^1 dz_q |\psi(z_q,r)|^2\\
\times
\int d^2z\frac{r^2}{z^2(r-z)^2}
({\mathcal S}(Y_\pom,z){\mathcal S}(Y_\pom,r-z)
-{\mathcal S}(Y_\pom,r))^2\ .
\label{final}
\end{multline}
$\sigma_0$ is a parameter, and two more parameters enter ${\mathcal S}$.
All of them are completely determined by the GBW fit to the $F_2$ structure
function \cite{GBW}.

The agreement with the data so far published by the ZEUS collaboration \cite{zeus}
is good, see Fig.1.
We expect that much could be learned on the saturation regime of QCD
from observables such as $d\sigma_{diss}/d\log Q^2$ if that could
be measured in the same kinematical regime \cite{GayDucati:2000ew}.
We refer the reader to Ref.\cite{Munier:2003zb} for further details.

\begin{figure}
\begin{center}
\epsfig{file=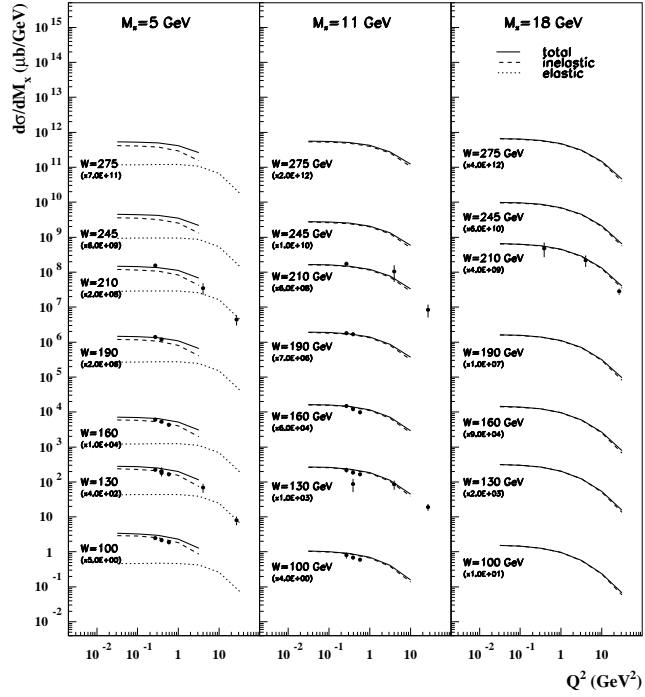,width=10.0cm}
\end{center}
\caption{Comparison of the model with the data. 
Dashed line: Eq.(\ref{final}).
Dotted line: elastic component, taken from GBW.
Full line: total. Data points are from the ZEUS coll.}
\end{figure}





\begin{thebibliography}{0}

\bibitem{Munier:2003zb}
S.~Munier and A.~Shoshi,
Phys.\ Rev.\ D {\bf 69}, 074022 (2004).

\bibitem{Good:1960ba}
M.~L. Good and W.~D. Walker,
Phys. Rev. {\bf 120}, 1857 (1960).


\bibitem{Mueller:1993rr}
A.~H.~Mueller,
Nucl.\ Phys.\ B {\bf 415}, 373 (1994).

\bibitem{Balitsky:1996ub}
I.~Balitsky,
Nucl. Phys. {\bf B463}, 99 (1996);
Y.~V. Kovchegov,
Phys. Rev. {\bf D60}, 034008 (1999).

\bibitem{Munier:2003vc}
S.~Munier and R.~Peschanski,
Phys.\ Rev.\ Lett.\  {\bf 91}, 232001 (2003);
Phys.\ Rev.\ D {\bf 69}, 034008 (2004);
arXiv:hep-ph/0401215.


\bibitem{kl}
Y.~V.~Kovchegov and E.~Levin,
Nucl.\ Phys.\ B {\bf 577}, 221 (2000).

\bibitem{Kovchegov:2001ni}
Y.~V.~Kovchegov,
Phys.\ Rev.\ D {\bf 64} (2001) 114016
[Erratum-ibid.\ D {\bf 68} (2003) 039901].


\bibitem{GBW}
K.~Golec-Biernat and M.~Wusthoff,
 Phys. Rev. {\bf D59}, 014017 (1999);
{\it ibid.} {\bf D60}, 114023 (1999).

                                                                                
\bibitem{zeus}
ZEUS, S.~Chekanov {\em et~al.},
 Eur. Phys. J. {\bf C25}, 169 (2002); see also
the talk by M. Ruspa, these proceedings.

                                                                                



\bibitem{GayDucati:2000ew}
M.~B.~Gay Ducati, V.~P.~B.~Goncalves and M.~V.~T.~Machado,
Phys.\ Lett.\ B {\bf 506}, 52 (2001); {\it idem},
Nucl.\ Phys.\ A {\bf 697}, 767 (2002).
                                                                                \end{thebibliography}
\end{document}